\documentclass[preprint,showpacs,preprintnumbers,amsmath,amssymb, showkeys,superscriptaddress,titlepage]{revtex4-2}
\usepackage{graphicx}
\usepackage{dcolumn}
\usepackage{bm}
\usepackage{hyperref}
\usepackage{textcomp}

\begin{document}

\title{Study of the Involvement of $^8$Be and $^9$B Nuclei in the Dissociation of Relativistic $^{10}$C, $^{10}$B, and $^{12}$C Nuclei}

\author{D.A. Artemenkov}
\affiliation{Joint Institute for Nuclear Research (JINR), Dubna, Russia}

\author{V. Bradnova}
\affiliation{Joint Institute for Nuclear Research (JINR), Dubna, Russia}

\author{G.I. Britvich}
\affiliation{Institute for High Energy Physics, National Research Center Kurchatov Institute, Protvino, Russia}

\author{A.A. Zaitsev}
\affiliation{Joint Institute for Nuclear Research (JINR), Dubna, Russia}
\affiliation{P.N. Lebedev Physical Institute of the Russian Academy of Sciences (LPI), Moscow, Russia}

\author{P.I. Zarubin}
\affiliation{Joint Institute for Nuclear Research (JINR), Dubna, Russia}
\affiliation{P.N. Lebedev Physical Institute of the Russian Academy of Sciences (LPI), Moscow, Russia}

\author{I.G. Zarubina}
\affiliation{Joint Institute for Nuclear Research (JINR), Dubna, Russia}

\author{V.A. Kalinin}
\affiliation{Institute for High Energy Physics, National Research Center Kurchatov Institute, Protvino, Russia}

\author{R.R. Kattabekov}
\affiliation{Joint Institute for Nuclear Research (JINR), Dubna, Russia}

\author{N.K. Kornegrutsa}
\affiliation{Joint Institute for Nuclear Research (JINR), Dubna, Russia}

\author{M.Yu. Kostin}
\affiliation{Institute for High Energy Physics, National Research Center Kurchatov Institute, Protvino, Russia}

\author{A.V. Maksimov}
\affiliation{Institute for High Energy Physics, National Research Center Kurchatov Institute, Protvino, Russia}

\author{K.Z. Mamatkulov}
\affiliation{Joint Institute for Nuclear Research (JINR), Dubna, Russia}

\author{E.K. Mitsova}
\affiliation{Joint Institute for Nuclear Research (JINR), Dubna, Russia}
\affiliation{South-Western University, Blagoevgrad, Bulgaria}

\author{A. Neagu}
\affiliation{Institute of Space Science, Magurele, Romania}

\author{V.A. Pikalov}
\affiliation{Institute for High Energy Physics, National Research Center Kurchatov Institute, Protvino, Russia}

\author{ M.K. Polkovnikov}
\affiliation{Institute for High Energy Physics, National Research Center Kurchatov Institute, Protvino, Russia}

\author{P. A. Rukoyatkin}
\affiliation{Institute of Space Science, Magurele, Romania}

\author{V.V. Rusakova}
\affiliation{Joint Institute for Nuclear Research (JINR), Dubna, Russia}

\author{V.R. Sarkisyan}
\affiliation{Yerevan Physics Institute, Yerevan, Armenia}

\author{R. Stanoeva}
\affiliation{South-Western University, Blagoevgrad, Bulgaria}

\author{M. Haiduc}
\affiliation{Institute of Space Science, Magurele, Romania}

\author{E. Firu}
\affiliation{Institute of Space Science, Magurele, Romania}

\author{S.P. Kharlamov}
\affiliation{P.N. Lebedev Physical Institute of the Russian Academy of Sciences (LPI), Moscow, Russia}

\begin{abstract}
The results obtained by estimating the contribution of $^8$Be and $^9$B nuclei to the coherent
dissociation of $^{10}$C, $^{10}$B, and $^{12}$C relativistic nuclei in nuclear track emulsions (``white'' stars) are presented.
The selection of ``white'' stars accompanied by $^9$B leads to a distinct peak appearing in the distribution of the
excitation energy of 2$\alpha$2$p$ ensembles and having a maximum at 4.1 $\pm$ 0.3 MeV. A $^8$Be nucleus manifests
itself in the coherent-dissociation reaction $^{10}$B $\to$ 2He + H with a probability of (25 $\pm$ 5)\%, (14 $\pm$ 3)\% of
it being due to $^9$B decays. The ratio of the branching fractions of the $^9$B + $n$ and $^9$Be + $p$ mirror channels
is estimated at 6 $\pm$ 1. An analysis of the relativistic dissociation of $^{12}$C nuclei in a nuclear track emulsion
revealed nine 3$\alpha$ events corresponding to the Hoyle state.
\end{abstract}

\maketitle

\section{Introduction}

A relativistic approach to studying nucleon clustering
on light nuclei was developed within the
BECQUEREL project \cite{1} (for an overview, see \cite{2,3}).
The project was aimed at an analysis of layers of
nuclear track emulsion that were exposed to primary
and secondary beams of nuclei accelerated to an
energy of about 1 GeV per nucleon at the Nuclotron
of the Joint Institute for Nuclear Research (JINR,
Dubna). The accelerator facility at the Institute for
High Energy Physics (IHEP, Protvino), which may
provide beams of $^{12}$C nuclei whose energy ranges
from a few hundred MeV units (booster) to several
tens of GeV units (main ring), opens new possibilities
in this respect. In this context, we present below the
results and prospects of studies of multiparticle states
involving $^8$Be and $^9$B unstable nuclei originating from
the dissociation of a $^{12}$C radioactive nucleus and $^{10}$B
and $^{12}$C stable nuclei.

Observations of events of the multiparticle fragmentation
of relativistic nuclei by means of nuclear
track emulsions are unique in completeness and
in angular resolution. In order to deduce conclusions
on the structure of nuclei being studied,
one analyzes the observed interactions, focusing on
coherent-dissociation events, which do not involve
slow fragments or charged mesons (``white'' stars).
This event selection gives grounds to assume that
respective collisions have a tangential character and
that colliding nuclei are minimally perturbed. White
stars emerge upon nuclear diffractive dissociation
without an overlap of the densities of colliding nuclei.
The probability for final states of fragments in ``white''
stars provides an estimate of their contribution to the
structure of nuclei being studied.

Unstable nuclei of $^8$Be and $^9$B may play a key role
in a general picture of nuclear clustering. Although
attempts at respective observations run into serious
difficulties, the contribution of these nuclei deserves
investigation over the whole available range of nuclei.
On the basis of measuring emission angles for
helium and hydrogen isotopes, one reconstructs the
decays of $^8$Be and $^9$B. The relativistic decays of $^8$Be
and $^9$B nuclei can be identified in the distributions of
the variable $Q = M^* - M$, where $M^{*2} = \Sigma(P_i \cdot P_k)$,
$M^*$ is the invariant mass of the system of fragments,
and $P_{i,k}$ stands for their 4-momenta determined in
the approximation where the fragments conserve the
primary momentum per nucleon.

\section{POSSIBLE 2$\alpha$2$\textbf{p}$ RESONANCE}
The structure of the $^{10}$C radioactive nucleus was
studied by the method of coherent dissociation at an
energy of 1.2 GeV per nucleon \cite{4}. It was found
that events of the 2He + 2H channel saturate 82\% of
``white'' stars. The assumption that helium nuclei correspond
to $^4$He, while hydrogen isotopes correspond
to proton is justified for $^{10}$C $\to$ 2He+2H ``white'' stars.
In analyzing the distributions of 2$\alpha p$ three-particle
combinations with respect to the energy $Q_{2\alpha}$, it was
found that $^9$B manifests itself in $^{10}$C with a probability
of (30 $\pm$ 4)\%, while $^8$Be$_{g.s.}$ originates from $^9$B decays
exclusively.

A feature missed in \cite{4} was found recently in
the energy ($Q_{2\alpha2p}$) distribution of 2$\alpha$2$p$ four-particle
combinations (Fig. 1a). This is a distinct peak (RMS
is 2.0 MeV) at $Q_{2\alpha2p}$ = 4.1 $\pm$ 0.3 MeV for white stars
featuring $^9$B decays. The number of events forming
this peak is (17 $\pm$ 4)\% of the total number of $^{10}$C ``white''
stars and (65 $\pm$ 14)\% of events involving $^9$B decay.
The distribution of all 2$\alpha$2$p$ ensembles with respect to
the total transverse momentum $P_{\textrm{T}2\alpha2p}$ (see Fig. 1b)
is described by the Rayleigh function specified by
the parameter value of $\sigma$ = 175 $\pm$ 10 MeV/$c$. In
the presence of $^9$B, this distribution is substantially
narrower - $\sigma$ = 127 $\pm$ 16 MeV/$c$.

\begin{figure}[th]
	\centerline{\includegraphics[width=15cm]{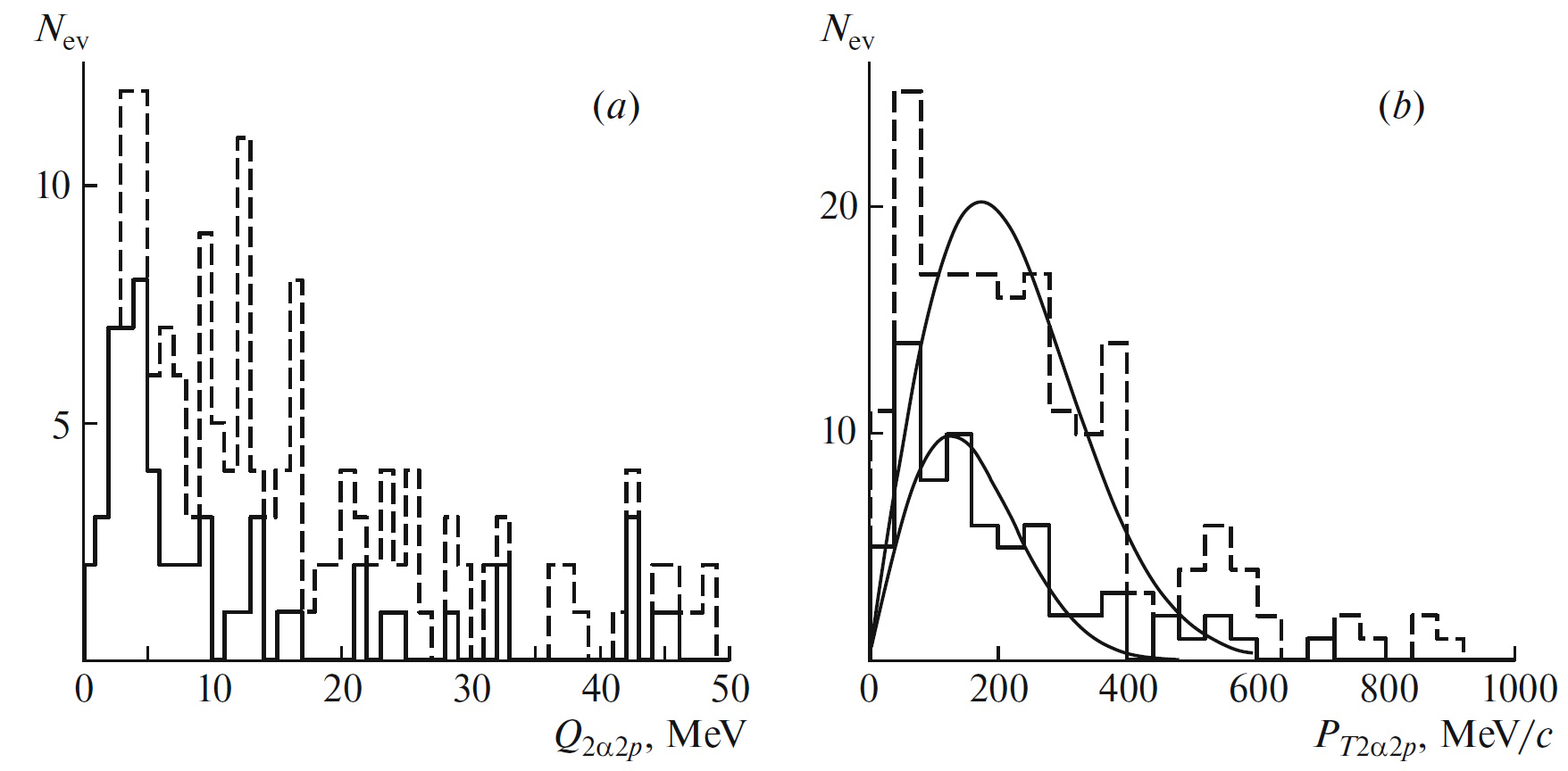}}
	\caption{(a) Energy ($Q_{2\alpha2p}$) and (b) total-transverse-momentum ($P_{\textrm{T}2\alpha2p}$) distributions of (dashed-line histogram) all $^{10}$C $\to$
		2He + 2H white stars and (solid-line histogram) stars that involve $^9$B.}
\end{figure}

An indication that such a resonance is present in
the $^9$B + $p$ system was obtained in \cite{5} by employing
$^{10}$C nuclei at the energy of 35 MeV per nucleon, but
this was not confirmed in a different experiment \cite{6}
where the energy of these nuclei was 10 MeV per nucleon.
Later, the authors of \cite{5} disavowed their original
result, referring to insufficient resolution and supporting
their statement by a simulation \cite{7}. Nevertheless,
a strong energy dependence of the inevitably
peripheral excitation of this resonance may underlie
the contradiction between \cite{5} and \cite{6}.

In low-energy experiments involving the detection
of all projectile fragments, the condition of its
peripheral character cannot be strengthened by the
requirement of the absence of target fragments. At
the energy threshold for the dissociations reaction,
such a resonance may either prove to be unobservable
under conditions of an intricate reaction mechanism
or not arise in principle. Our observation is based
on a totally different implementation. It is maximally
reliable and boasts the highest angular resolutions
in measuring tracks of 2$\alpha$2$p$ four-particle combinations.
In order to confirm the existence of such a
resonance, which may prove to be a 2$\alpha$2$p$ nuclear-molecule
system, it is highly desirable to obtain a
vaster data sample on the basis of a new irradiation
run and to apply a faster method of searches for jets of
2He + 2H fragments.

\section{ASYMMETRY IN MIRROR CHANNELS OF DISSOCIATION OF $^{10}$B}
An irradiation of a nuclear track emulsion with
$^{10}$B nuclei accelerated to an energy of 1 GeV per
nucleon was implemented in one of the first runs
at the JINR Nuclotron. An analysis of coherent dissociation
events revealed the dominance of ``white''
stars in the 2He + H channel and their suppression
in the Be + H channel (its fraction was not more than
2\%). That the contribution of the $^9$B unstable nucleus
to the structure of the $^{10}$C radioactive nucleus
was found to be significant \cite{4} was indicative of the
possible involvement of $^9$B in the dissociation of $^{10}$B.
Moreover, an interpretation of the structure of the
$^{10}$B nucleus with allowance for the possible presence
in it of a superposition of $^8$Be$_{g.s.}$ and $^8$Be$_{2^+}$ virtual
states becomes possible upon resorting to appearing
information about the dissociation of $^9$Be nuclei in
a nuclear track emulsion. A determination of the
contribution of unstable nuclei furnishes the basis for
obtaining deeper insight into the structure of $^{11}$C \cite{3}
and $^{12}$N nuclei.

The repeated scanning in 2015 along the tracks
of $^{10}$B beam nuclei over the length of 241 m resulted
in finding 1664 nuclear stars \cite{3}. The distribution
of 127 $^{10}$B white stars found among them confirms
the dominance of the 2He + H channel (78\%)
and the suppression of the Be + H channel (1\%),
which should correspond to the $^9$Be + $p$ configuration.
The remaining events were distributed among
the He + 3H (12\%), Li + He (4\%), and Li + He (4\%)
channels.

In the measurements of the divergence angle
$\Theta_{2\textrm{He}}$, the sample of 2He pairs in the range of 0 $<$
$\Theta_{2\textrm{He}}$ $<$ 10.5 mrad, which corresponds to $^8$Be$_{g.s.}$
decay, includes (25 $\pm$ 5)\% of $^{10}$B $\to$ 2He + H white
stars (see Fig. 2a). The divergence-angle interval
of $\Theta$($^8$Be$_{g.s.}$ + H) $<$ 25 mrad (see Fig. 2b), which
corresponds to $^9$B decays, contains only (14 $\pm$ 3)\% of
$^{10}$B $\to$ 2He + H white stars. Thus, $^9$B decays explain
only (5 $\pm$ 16)\% of $^8$Be$_{g.s.}$ decays. It should be emphasized
that this conclusion differs from that for the case
of $^{10}$C, where there was a perfect correspondence.
This lends support to the assumption that, in the
$^{10}$B nucleus, there is a superposition of $^8$Be$_{g.s.}$/$^8$Be$_{2^+}$
cores, along with $^9$B. This superposition manifests
itself as an excess of events involving $^8$Be$_{g.s.}$.

\begin{figure}[th]
	\centerline{\includegraphics[width=15cm]{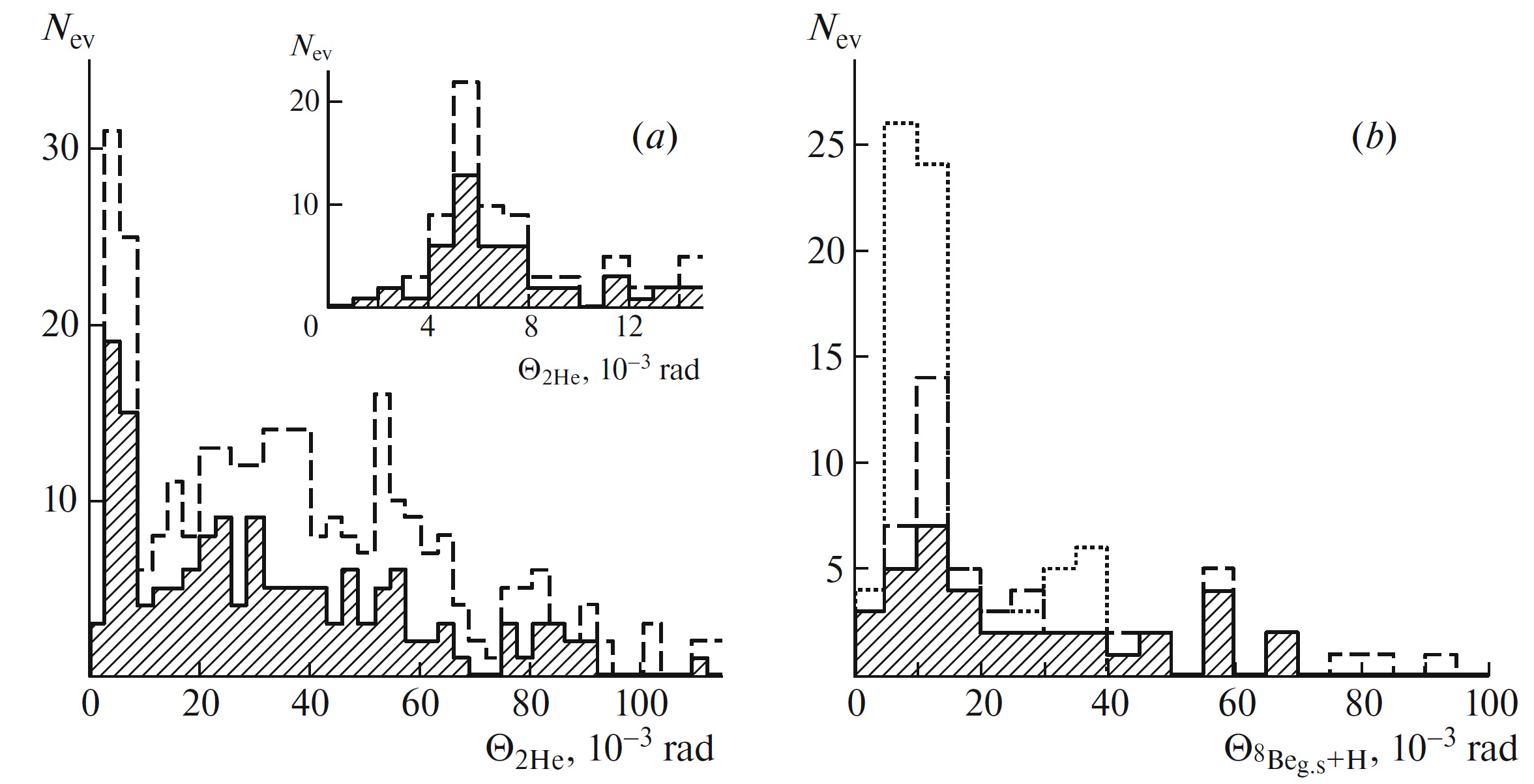}}
	\caption{Distribution of (a) the angle of divergence, $\Theta_{2\textrm{He}}$, of helium fragments in (dashed-line histogram) all stars and (shaded
		region) white stars involving $^{10}$B and (b) the angle of divergence, $\Theta$($^8$Be$_{g.s.}$ + H), in (dotted-line histogram) white stars
		involving $^{10}$C, (dashed-line histogram) all stars, and (shaded region) white stars involving $^{10}$B.}
\end{figure}

The set of $^{10}$B white stars found without selections
permits estimating the ratio of the branching fractions
of the $^9$B + $n$ and $^9$Be + $p$ channels at 6 $\pm$ 1, which
seems unexpected. An alternative explanation of the
asymmetry of the branching fractions of these mirror
channels could be based on a qualitatively broader
distribution of neutrons in relation to protons. However,
this version seems improbable, since the inelastic
cross sections for the interaction of relativistic $^{10}$B
nuclei do not exhibit exotic behavior.

Possibly, this fact is indicative of the presence of
the $^9$Be core in $^{10}$B, predominantly in the form of a
$^8$Be$_{2^+}$/$^8$Be$_{g.s.}$ + $n$ superposition (nuclear molecule),
manifesting itself in the dissociation mode involving
the direct formation of $^8$Be$_{g.s.}$ (without $^9$B decays) and
$^8$Be$_{2^+}$ states. As for the $^9$B core, it initially has this
cluster form.

Relativistic hydrogen and helium fragments can be
identified by the parameter $p\beta c$, which is determined
on the basis of measurements of multiple scattering
of tracks in a nuclear track emulsion, where $p$ is the
total momentum and $\beta c$ is the speed. Measurement
of the parameter $p\beta c$ for beam deuterons (calibration)
makes it possible to test this cumbersome and
not always implementable procedure (see Fig. 3).
For their 20 tracks, the average value of $\left\langle p\beta c \right\rangle$  was
2.5 $\pm$ 0.5 GeV at RMS = 0.6 GeV; this complies with
the expected value. In addition, a value of $\left\langle p\beta c \right\rangle$ =
1.1 $\pm$ 0.3 GeV at RMS = 0.4 GeV was obtained
for hydrogen fragments from $^9$B decays in $^{10}$B $\to$
2He + H white stars. This corresponds both to decay
protons and to the primary momentum of $^{10}$B nuclei
per nucleon.

\begin{figure}[th]
	\centerline{\includegraphics[width=6cm]{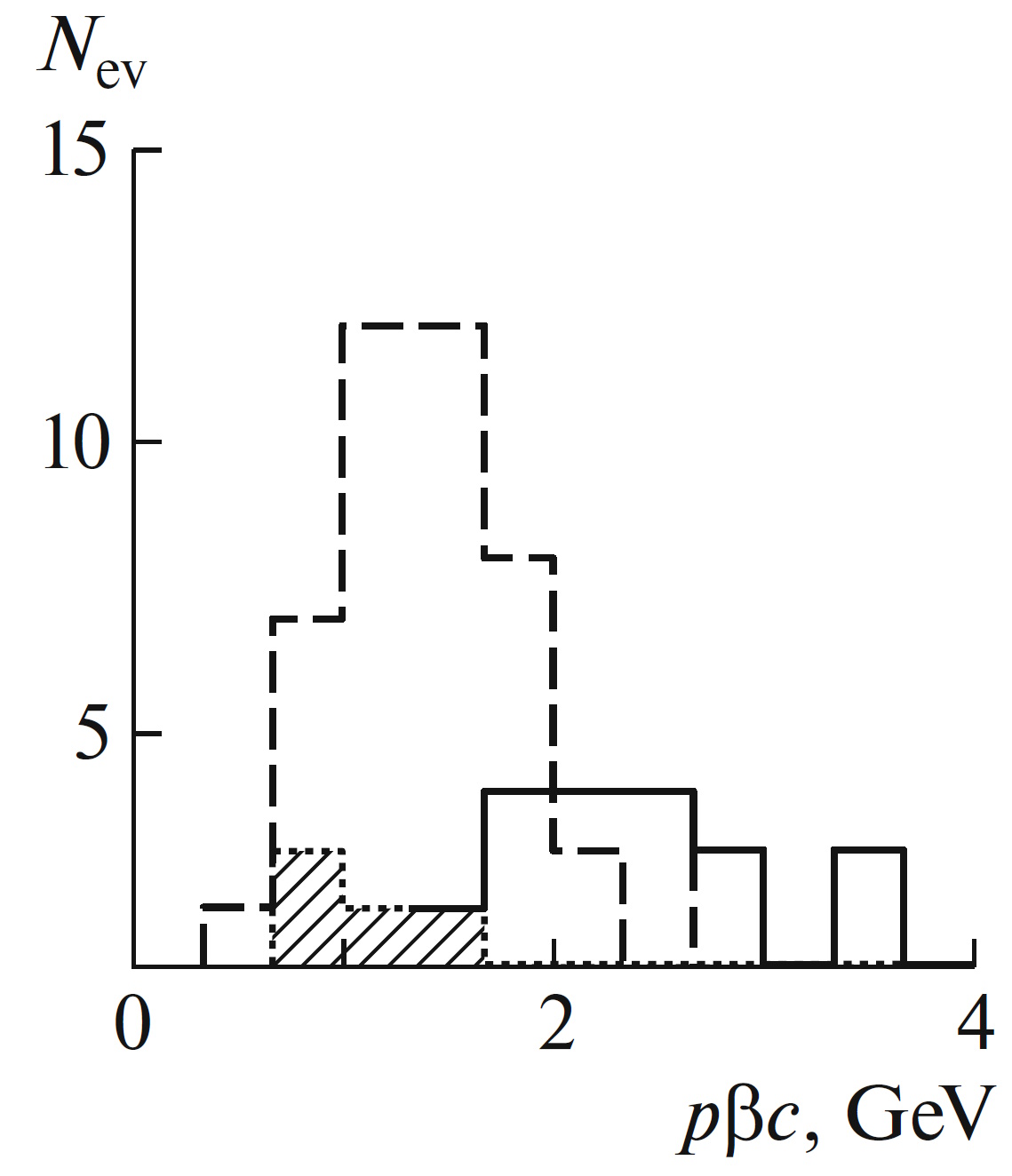}}
	\caption{$p\beta c$ distribution of tracks of (solid-line histogram)
		beam deuterons and (dashed-line histogram) hydrogen
		fragments in 52 $^{10}$B $\to$ 2He + H stars, including 36
		white stars and nine $^9$B decays (shaded region).}
\end{figure}

In order to identify hydrogen isotopes in $^{10}$B $\to$
2He + H white stars, we employed the procedure developed
on the basis of constructing classifying functions.
These functions were obtained from a statistical
simulation of the average values of the second differences
of track deviations $\left\langle |D| \right\rangle $ in cells of dimension
500, 600, 700, and 800 $\mu$m. The functions obtained
in this way were used to associate the experimental
values of $\left\langle |D| \right\rangle $ with groups of values characteristic of
different isotopes.

According to this procedure, protons characterized
by an average value of $\left\langle p\beta c \right\rangle$ = 1.2 $\pm$ 0.1 GeV
at RMS = 0.3 GeV dominate the $p\beta c$ spectrum of
hydrogen fragments in $^{10}$B $\to$ 2He + H white stars
up to 1.8 GeV. The $p\beta c$ distribution above 1.8 GeV
corresponds to deuterons for which $\left\langle p\beta c \right\rangle$ = 2.5 $\pm$
0.5 GeV at RMS = 0.7 GeV. The ratio of the numbers
of identified $p$ and $d$ tracks is 2 $\pm$ 0.25.

The identification of helium and hydrogen isotopes
makes it possible to perform a more profound
analysis of $^{10}$B $\to$ 2He + H white stars. In particular,
the $^8$Be$_{2^+}$ + $d$ configuration may be a source
of $^8$Be$_{2^+}$ decays. In the present study, we observed
six white stars of this type. Figure 4a shows the
distribution of the opening angle $\Theta_{2\textrm{He}}$ for helium
fragments under the condition that $p\beta c$ is greater
than 1.9 GeV. The distribution in question is characterized
by a mean value of $\left\langle \Theta_{2\textrm{He}} \right\rangle$  = 30 $\pm$ 3 mrad
at RMS = 7 mrad. This corresponds to $\left\langle Q_{2\alpha} \right\rangle $ =
2.8 $\pm$ 0.5 MeV at RMS = 0.9 MeV and, hence, to
the $^8$Be$_{2^+}$ state. Figure 4b gives the distribution of
the energy $Q_{2\alpha d}$ for these six $^8$Be$_{2^+}$ + $d$ ensembles;
it has $\left\langle Q_{2\alpha d} \right\rangle $ = 4.1 $\pm$ 0.3 MeV at RMS = 0.9 MeV.
Although this observation has a rather low statistical
significance, it is indicative of the possible formation
of states belonging to the nuclear-molecule type.
Moreover, white stars in the channel $^{10}$B $\to$ $^6$Li + $\alpha$
are observed with a probability of 8\%. Therefore, one
can expect a contribution of the cluster structure of
$^6$Li to the 2$\alpha$ + $p$($d$) channel. However, investigation
along these lines would require a vaster set of ``white'' star
data.

\begin{figure}[th]
	\centerline{\includegraphics[width=10cm]{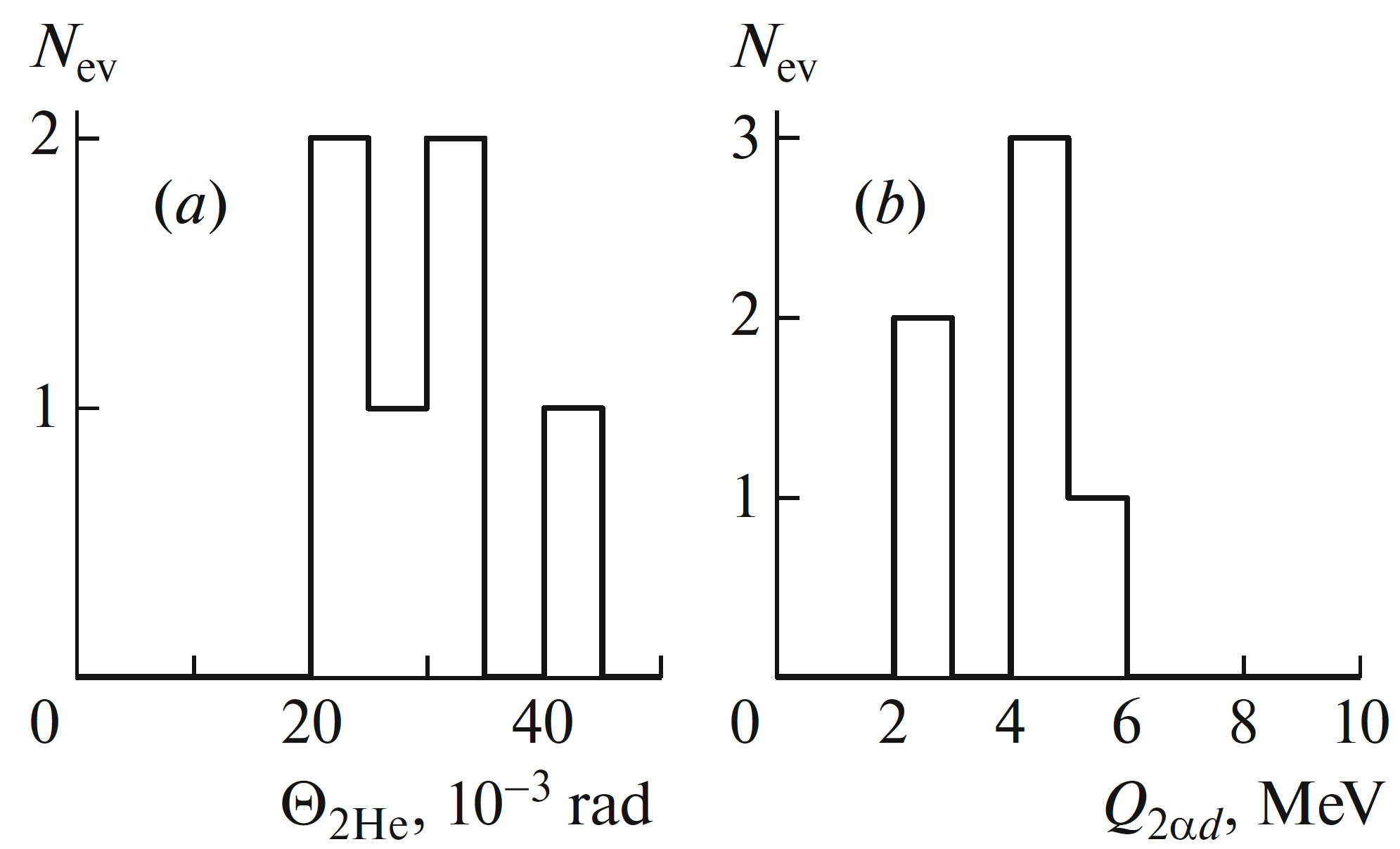}}
	\caption{Distribution of (a) the opening angle $\Theta_{2\textrm{He}}$ for
		helium fragments under the condition that $p\beta c$ is greater
		than 1.9 GeV (deuterons) in $^{10}$B $\to$ 2He +H white stars
		and (b) the energy $Q_{2\alpha d}$ for $^8$Be$_{2^+}$ + $d$ ensembles.}
\end{figure}

\section{SEARCH FOR THE HOYLE STATE}
A search for combinations of three alpha particles
in a Hoyle state (in the second excited state and
in the first unbound state of spin–parity 0$^+_2$ ) upon
the coherent dissociation of $^{12}$C nuclei is promising
task for the emulsion procedure. In nucleosynthesis,
the merger of a 3$\alpha$-particle ensemble into the $^{10}$C
nucleus proceeds through this state. It is assumed
that (along with $^8$Be) it corresponds to a Bose–
Einstein condensate \cite{8} of alpha particles that have
zero relative angular momenta. The Hoyle state is
expected to manifest itself experimentally in alpha-particle
jets that have extremely small opening angles
and in which all alpha-particle pairs correspond to
$^8$Be$_{g.s.}$. We will now expound on the experimental
evidence in support of this statement.

In the early 1970s, a nuclear track emulsion was
exposed to $^{12}$C nuclei of energy 3.65 GeV per nucleon
at the JINR synchrophasotron. Emulsion layers
stacks impregnated with lead salts were irradiated
later. We performed a reanalysis of these
irradiated emulsions and found nine events in which
the total transverse energy of alpha particles in the
reference frame comoving with the nucleus undergoing
dissociation corresponds to the excitation of the
$^{12}$C$^*$ to the level at 7.65 MeV (Hoyle state). Figure 5
gives the distribution of these events with respect to
the variable $Q_{3\alpha}$. These observations of priority character
motivate further searches for the Hoyle state at
acceptable statistical-significance level.

\begin{figure}[th]
	\centerline{\includegraphics[width=8cm]{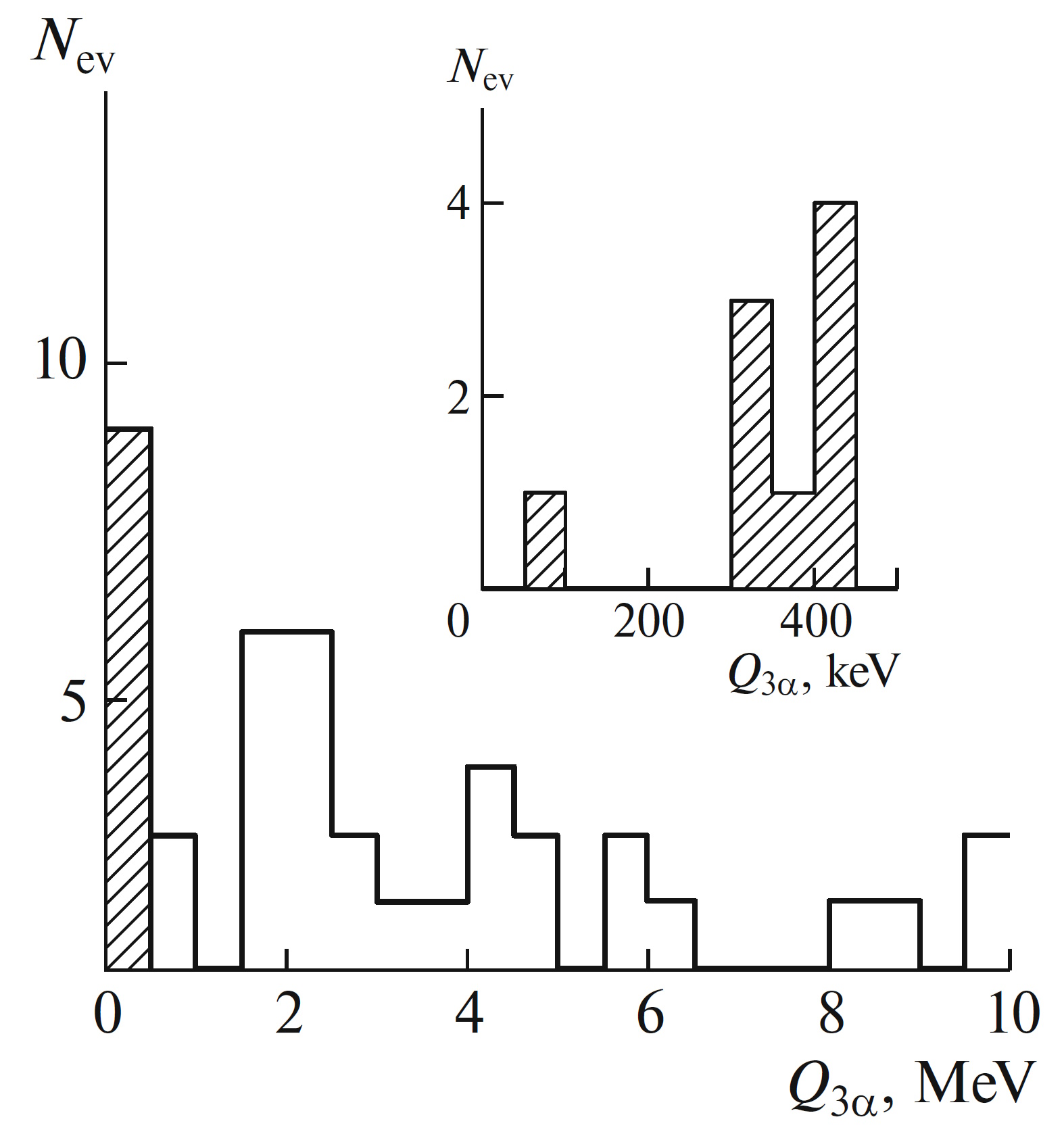}}
	\caption{Distribution of the energy $Q_{3\alpha}$ for combinations of
		three alpha particles in $^{12}$C $\to$ 3$\alpha$ coherent-dissociation
		events at a momentum of 4.5 GeV/$c$ per nucleon. The
		region between 0 and 500 keV is shown in the inset on a
		magnified scale.}
\end{figure}

For this purpose, emulsion stacks will be exposed
longitudinally to $^{12}$C nuclei at the JINR Nuclotron
and at the IHEP accelerator facility. The energy of
$^{12}$C nuclei will range from several hundred MeV units
to a few tens of GeV units per nucleon, and this will
permit tracing the behavior of the cross section for
the emergence of 3$\alpha$ combinations in the Hoyle state
and establishing the universality of its formation. The
nuclear track emulsion used will be enriched in lead
nuclei in order to enhance the contribution of electromagnetic
dissociation. We will measure opening
angles in several hundreds of combinations of three
alpha particles.

It is important to perform respective measurements
over broader regions of energies of $^{12}$C nuclei
since theoretical calculations of cross sections for the
electromagnetic dissociation of light nuclei predict a
broad maximum in the region of several hundred MeV
units per nucleon. At lower projectile energies, the
implementation becomes more convenient in several
aspects of practical importance. First, a visual contrast
between $\alpha$-pair fragments and narrow pairs from
$^8$Be$_{g.s.}$ decays becomes sharper. Second, the fraction
of background events involving the production of
charged mesons becomes lower. Third, it is easier
to level out the beam profile at the entrance of the
emulsion stack. The effect of deceleration of a primary
nucleus in the emulsion medium can be compensated
by the invariant representation of the $\alpha$-pair energy
and by the narrowing the scanned region along the
beam direction.

A medical beam of $^{12}$C nuclei at IHEP is used
at the initial and the mane stage of this project. It
ensures the required uniformity of irradiation and has
an energy that corresponds to the maximum of the
cross section for electromagnetic dissociation. Its
working intensity, which is not less than 10$^8$ nuclei
per cycle, should be reduced by a factor of not less
than 1000 in order to avoid excessive irradiation and
to accomplish beam monitoring. This presents a difficult
challenge since a high beam intensity provides
feedback for tuning the accelerator.

Work performed in December 2016 made it possible
to irradiate controllably track-emulsion layers
500 $\mu$m in thickness. The production of such layers
was renewed by the Slavich enterprise (Pereslavl-
Zalessky). In order to ensure a particle density of
2000 to 4500 nuclei/cm$^2$ at the irradiation locus,
we changed the slow-extraction mode, reduced the
extraction efficiency, additionally shifted the point of
emulsion exposure by 8 m along the beam direction,
and reduced the extraction time from 600 to 400 ms.

The irradiation of track emulsions was monitored
by means of three counters based on scintillators
produced at IHEP (polystyrene plastic of the STs-301
type). The counters, equipped with photomultiplier
tubes (PMT-85), were 1 cm thick and had cross-sectional
dimensions of 10 $\times$ 10 mm. The irradiated
emulsion stacks were placed in front of the counters.
Figure 6 shows the content of the beam of carbon
nuclei from the medical channel of the U-70 accelerator
at the locus of emulsion exposure. The beam
contained nuclei of charge number 6 (about 78\%), 5
(2\%), 4 (2\%), 3 (2\%), 2 (14\%), and 1 (2\%). This mixture
is an expected consequence of the absence of a
vacuum ion guide and separating magnets. This fact
does not complicate the selection of white stars. On
the contrary, there arises the possibility of additional
calibrations in charge and in multiple scattering. At
the present time, the irradiated samples are under
analysis.

\begin{figure}[th]
	\centerline{\includegraphics[width=10cm]{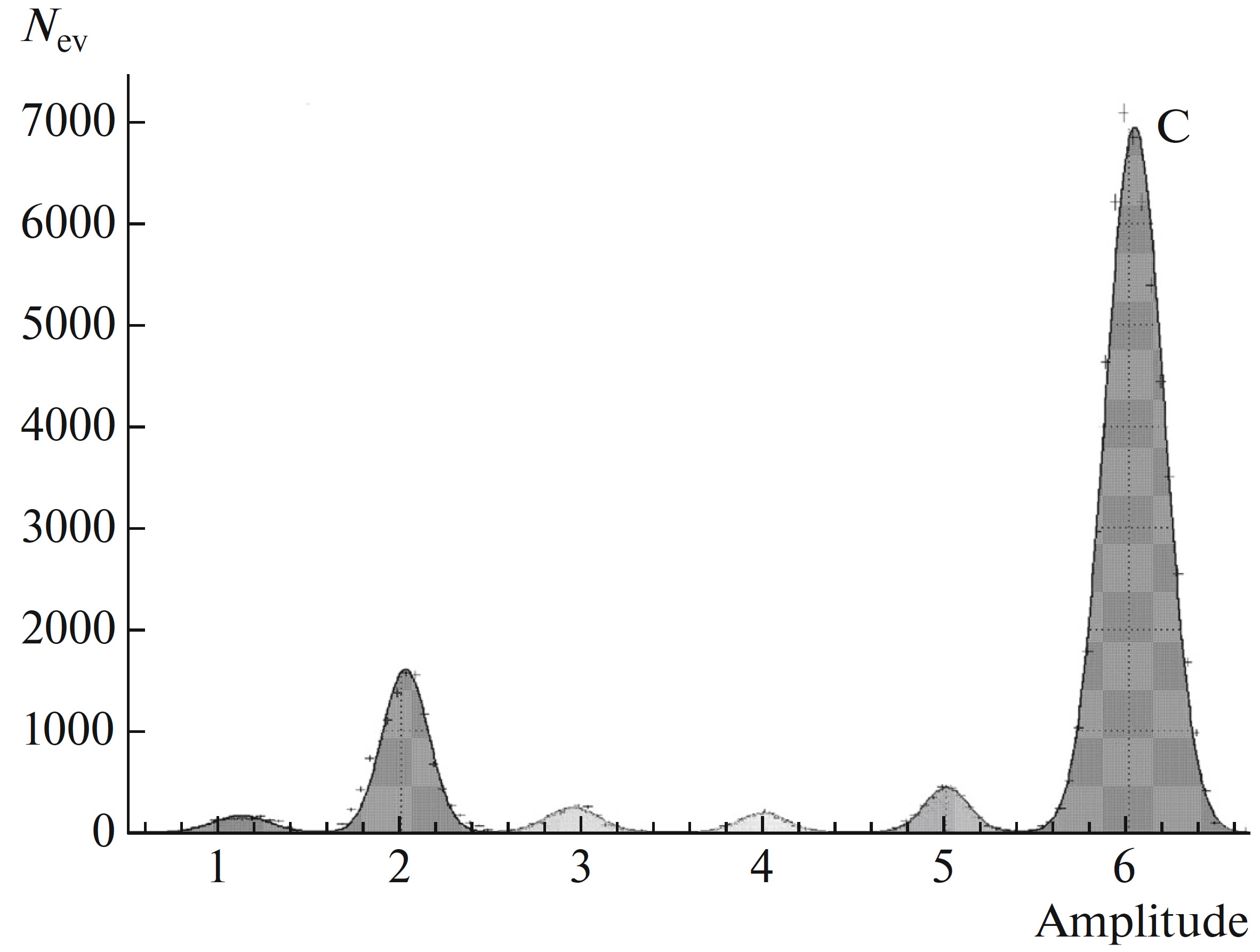}}
	\caption{Content of the beam of carbon nuclei from the medical channel of the U-70 accelerator at locus of emulsion exposure.}
\end{figure}

\section{CONCLUSIONS}

By and large, the family of light nuclei, which
form the beginning of the isotope table, remains a
``laboratory'' full of surprises, where one can trace
the emergence of the shell structure. The relativistic
dissociation of these nuclei, including radioactive
ones, in nuclear track emulsions, makes it possible
to study the whole variety of ensembles of extremely
light clusters up to the binding threshold. The emerging
physical picture can be summarized as follows.

In addition to nucleons, the structure of light nuclei
involves clusters, such as alpha particles, tritons,
$^3$He nuclei (helions), and deuterons, that do not have
excited states. An alpha-particle pair may form an
unstable nucleus of $^8$Be in the ground state and in the
first excited state: ($^8$Be$_{g.s.}$ and $^8$Be$_{2^+}$). These configurations
are present with nearly identical probabilities
as cores in the $^9$Be nucleus, which is stable. The $^7$Be
and $^7$Li stable nuclei play an important role in the
structure of heavier neutron-deficient and neutron-rich
nuclei. The $^9$B unstable molecule-like nucleus
($^8$Be$_{g.s.}$ + $p$) and the $^9$Be stable nucleus may play
nearly the same role as cores of the $^{10}$B nucleus. At
the same time, the $^9$Be nucleus may be present in
the $^{10}$B nucleus not only as a discrete unit but also
as a $^8$Be$_{g.s.}$/$^8$Be$_{2^+}$ + $n$ superposition similar to the
$^9$B($^8$Be$_{g.s.}$ + $n$) unbound nucleus.

A balanced coexistence of possible superpositions
of core nuclei, clusters, and nucleons determines
ground-state parameters of the respective nucleus.
Despite the relativistic scale of coherent interactions,
searches for effects of few-body quantum mechanics
and even for manifestations of nuclear-molecule
systems can be performed in them. However, only
the track-emulsion method possesses the required
spatial resolution.

The absence of $^8$Be and $^9$B bound nuclei among
existing isotopes has a crucial effect on the propagation
of nucleosynthesis, making it circumvent these
gaps in the isotope table. The absence of stable
ground states of these nuclei does not prevent their
involvement in the nuclear structure as cores. An
analysis of the nucleosynthesis path to $^{10,11}$B through
the chain $^7$Be($^3$He, $\gamma$)$^{10}$C(e$^+$, $\nu$)$^{10}$B (hot breakout) provides an estimate of their importance.

The synthesis of $^{10}$C is ensured by an energy ``window''
for the formation of the $^9$B + $p$, $^8$Be$_{2^+}$ + 2$p$, and
$^6$Be + $\alpha$ intermediate states. These configurations
survive in the subsequent chain
$^{10}$C(e$^+$, $\nu$)$^{10}$B($p$,$\gamma$)$^{11}$C(e$^+$, $\nu$)$^{11}$B. The ``window'' for
the reaction $^7$Be($^4$He,$\gamma$)$^{11}$C permits only the fusion of
$^7$Be and $^4$He, which also contributes to the structure
of $^{11}$C and $^{11}$B.

Thus, the population of the hidden variety of virtual
configurations in the $^{10,11}$C and $^{10,11}$B nuclei proceeds
through electromagnetic transitions from real
states involving unstable nuclei. In turn, these nuclei
participate in the synthesis of subsequent nuclei
via their involvement in proton-capture (or neutron-exchange)
reactions or reactions of the addition of
helium isotopes, and this leads to inheriting special
features of preceding structures.

The identification of the Hoyle state in the dissociation
of $^{12}$C would open prospects for searches by
this method in the dissociations of heavier nuclei for
condensate states featuring a large number of alpha
particles. In turn, the discovery of alpha-particle condensates
would enable one to consider new scenarios
of nuclear astrophysics.

\end{document}